\documentclass[twocolumn,prb,superscriptaddress]{revtex4}
\AtBeginDocument{
\heavyrulewidth=.08em
\lightrulewidth=.05em
\cmidrulewidth=.03em
\belowrulesep=.65ex
\belowbottomsep=0pt
\aboverulesep=.4ex
\abovetopsep=0pt
\cmidrulesep=\doublerulesep
\cmidrulekern=.5em
\defaultaddspace=.5em
}

\usepackage{amsmath,amssymb}
\usepackage{graphicx}
\usepackage[usenames,dvipsnames]{xcolor}
\usepackage{amsthm}
\usepackage{subfigure}
\usepackage{booktabs}
\usepackage{hyperref}
\usepackage{accents}
\newcommand{\ubar}[1]{\underaccent{\bar}{#1}}
\usepackage{bbold}
\bibpunct{[}{]}{,}{n}{}{}
\usepackage[printwatermark]{xwatermark}
\hypersetup{colorlinks=true,citecolor=Red,linkcolor=Blue,urlcolor=Black}
\usepackage[export]{adjustbox}

\begin{document}

\title {Doping evolution of charge and spin excitations in two-leg Hubbard ladders: 
comparing DMRG and RPA+FLEX results}

\author{A. Nocera}
\affiliation{Department of Physics and Astronomy, University of Tennessee, Knoxville, 
Tennessee 37996, USA}
\affiliation{Materials Science and Technology Division, Oak Ridge National Laboratory, Oak Ridge, Tennessee 37831, USA}

\author{Y. Wang}
\affiliation{Department of Physics and Astronomy, University of Tennessee, Knoxville, 
Tennessee 37996, USA}
\affiliation{D{\'e}partement de Physique and Institut Quantique, Universit{\'e} de Sherbrooke, Sherbrooke, Qu{\'e}bec J1K 2R1, Canada}

\author{N. D. Patel}
\affiliation{Department of Physics and Astronomy, University of Tennessee, Knoxville, 
Tennessee 37996, USA}
\affiliation{Materials Science and Technology Division, Oak Ridge National Laboratory, Oak Ridge, Tennessee 37831, USA}

\author{G. Alvarez}
\affiliation{Computational Science and Engineering Division and Center for Nanophase Materials Sciences, Oak Ridge National Laboratory, Oak Ridge, Tennessee 37831, USA}

\author{T. A. Maier}
\affiliation{Computational Science and Engineering Division and Center for Nanophase Materials Sciences, Oak Ridge National Laboratory, Oak Ridge, Tennessee 37831, USA}

\author{E. Dagotto}
\affiliation{Department of Physics and Astronomy, University of Tennessee, Knoxville, 
Tennessee 37996, USA}
\affiliation{Materials Science and Technology Division, Oak Ridge National Laboratory, Oak Ridge, Tennessee 37831, USA}

\author{S. Johnston}
\affiliation{Department of Physics and Astronomy, University of Tennessee, Knoxville, Tennessee 37996, USA}
\affiliation{Joint Institute for Advanced Materials at The University of Tennessee, Knoxville, Tennessee 37996, USA}

\begin{abstract}
We study the magnetic and charge dynamical response of a Hubbard model in a two-leg ladder geometry using the density matrix renormalization group (DMRG) method and the random phase approximation within the fluctuation-exchange approximation (RPA+FLEX). 
Our calculations reveal that RPA+FLEX can capture the main features of the magnetic response from weak up to intermediate 
Hubbard repulsion for doped ladders, when compared with the numerically exact DMRG results. However, while at weak Hubbard repulsion both the spin and charge spectra 
can be understood in terms of weakly-interacting electron-hole 
excitations across the Fermi surface, at intermediate coupling DMRG shows gapped spin excitations at large momentum transfer that remain gapless within the RPA+FLEX approximation. For the charge response, RPA+FLEX can only reproduce the main features of the DMRG spectra at weak coupling and high doping levels, while it shows an incoherent character away from this limit.
Overall, our analysis shows that RPA+FLEX works surprisingly well for spin excitations at weak and intermediate 
Hubbard $U$ values even in the difficult low-dimensional geometry such as a two-leg ladder. Finally, 
we discuss the implications of our results for neutron scattering and resonant inelastic x-ray scattering experiments
on two-leg ladder cuprate compounds.
\end{abstract}

\maketitle

\section{Introduction}

Thirty years since the discovery of high critical temperature superconductivity in cuprates, 
understanding the microscopic mechanism leading to pairing remains a challenge. 
Progress on this problem has been hindered mainly by the lack of a viable numerical 
solution of the two dimensional Hubbard model~\cite{PhysRevX.5.041041}, 
which shows competition between different phases in the weakly hole doped regime, including $d$-wave superconductivity, pseudo-gap, 
and charge-density-wave (stripes) phases\cite{Dagotto257,Norman196,Comin2016}. The limited knowledge 
about the ground state of this model has made the study of its magnetic and charge excitations and their 
doping dependence even more challenging~\cite{white1989, jia2013persistent, kung2015, yang2016}.
In this context, a set of surprising experimental results have emerged from recent resonant inelastic x-ray 
scattering measurements (RIXS)~\cite{re:Ament2011,Comin2016}. In the hole-doped 
cuprate families~\cite{lipscombe2007,le2011intense,letacon2013,dean2013,dean2013persistence,lee2014asymmetry,re:Wakimoto2015},
high-energy magnons or paramagnons on the antiferromagnetic zone boundary persist from the 
parent compounds into the heavily overdoped regime, showing little doping dependence up 
to 40\% hole doping, where the system is believed to exhibit Fermi-liquid-like behavior. 
This observation is in contrast to neutron scattering experiments~\cite{coldea2001,wakimoto2004,wakimoto2005}, 
which find that the low-energy magnetic excitations gradually disappear around wavevector ${\bf q} = (\pi,\pi)/a$ 
with doping into the overdoped regime. These observations have shown that assessing 
the role of both the low- and high-energy magnetic excitations in the superconductivity of cuprates still deserves further attention. 

Because of these challenges, the study of quasi-one-dimensional (1D) cuprate systems such as two-leg ladders 
has become of interest as a simpler starting point for understanding the layered two-dimensional 
systems~\cite{re:Dagotto1992, dagotto1995surprises, dagotto1999experiments}. 
One of the reasons is that numerical calculations can be done more accurately for model Hamiltonians in 1D or quasi-1D systems. 
Indeed, different many-body techniques have successfully unveiled interesting properties of the Hubbard model in a two-leg ladder geometry 
such as an unusual spin gap in the undoped state~\cite{re:Barnes1993,Noack94}, 
and superconducting $d$-wave-like tendencies in the weakly doped regime~\cite{PhysRevB.92.195139}. 

Experiments have verified many theoretical predictions for these quasi-1D systems. 
For example, NMR~\cite{azuma1994,Kumagai1997,Imai1998} and neutron 
experiments~\cite{PhysRevB.88.014504} have observed a robust gap upon doping in the so-called 
``telephone number'' compound Sr$_{14-x}$Ca$_{x}$Cu$_{24}$O$_{41}$~\cite{Vuletic2006}, 
while superconductivity with a critical temperature of $T_\mathrm{c} = 12$~K has been reported in the same material under 
high pressure~\cite{uehara1996superconductivity, PhysRevLett.81.1090}. These results provide considerable 
support to the notion that superconductivity in cuprates in the weakly doped regime originates from antiferromagnetic fluctuations. 
The magnetic excitations of the ground state of the cuprate two-leg ladders have also been measured to a high degree
of accuracy in the undoped regime. Neutron scattering experiments have observed both one-triplon 
and two-triplon excitations~\cite{eccleston1996,notbohm2007}, 
which are the analog of magnon and bi-magnon excitations in the layered systems. 
Recent RIXS experiments have also successfully observed the two-triplon excitations~\cite{re:schlappa2009}. 

Much less is known about the cuprate two-leg ladders at high doping levels. 
In the layered systems, one expects that spin excitations behave like weakly interacting particle-hole 
excitations governed by the underlying free particle kinetic energy, with a minor influence from 
the Hubbard interaction $U$. If this notion is correct, then this high doping limit 
should be adequately described by the random phase approximation (RPA)~\cite{chen1991,kung2015}. 
Indeed, many studies have assumed weak 
correlations in doped cuprates in the 
layered geometry~\cite{scalapino2012, Bulut1993b, monthoux1994, scalapino1995case, Moriya2003, dahm1997temperature, maier2007spin, maier2007systematic}, 
and used the RPA to study the spin and charge excitations 
in comparison to neutron and Raman scattering experiments, as well as the formation of a $d$-wave superconducting state. 
 
In this context, quasi-1D systems provide an excellent opportunity to explore how both spin and charge excitations 
systematically evolve with doping throughout the Brillouin zone.
 These same systems also offer a means to assess the 
degree to which RPA can capture various response functions that be evaluated with exact numerical techniques such as density matrix renormalization 
group (DMRG)~\cite{re:white92,re:white93}.
With this motivation, in the present work we compute the dynamical spin and charge response functions of the single-band 
Hubbard model on a two-leg ladder geometry using DMRG~\cite{re:Kuhner1999,nocera2016spectral}. We then compare the spin and charge structure factors to those obtained with a fully 
self-consistent RPA formalism, in which the interacting Green's function is obtained within the fluctuation-exchange approximation (FLEX)~\cite{takimoto2004strong, Yada2005, Zhang2009, Zhang2010}. 
The RPA formalism~\cite{Bulut1992a,Bulut1993a,Bulut1993b} was initially developed for weakly interacting systems and is expected to become an increasingly good approximation as the doping level increases. 
FLEX has been applied to the case of the one-band Hubbard model 
for cuprates~\cite{st1994carbotte, monthoux1994self, dahm1995quasiparticle, langer1995theory, grabowski1996theory, altmann2000anisotropic,Kontani1998}, and has been generalized to the multiband case (see Refs.~\cite{manske2003renormalization, takimoto2004strong}). Our calculations reveal that, while RPA describes well the spin response from weak to intermediate values of the Hubbard $U$, it fails to reproduce the dispersion of the main features in the strong coupling regime. On the other hand, 
RPA can reproduce the charge response only at weak coupling and high doping. Nevertheless, RPA+FLEX works surprisingly well in the
spin sector up to an intermediate $U$ even in the more challenging low-dimensional geometry of a two-leg ladder where the correlation effects are larger due to a narrower bandwidth.

This work is organized as follows: Section~II and III introduce the model and the methods, respectively. Section~IV presents the main results. Section~IV.A explores the pairing symmetry in the ground state of the two-leg ladder system. Section~IV.B presents results for the charge and spin dynamical structure factors of the Hubbard two-leg ladder in the weak coupling regime. Sec.~IV.C and IV.D explore the excitation spectra in the intermediate and strong coupling regimes. Finally, Section~V provides a summary of the results with a sketch of the range of validity for the RPA approximation, a discussion about the implications of our results for neutron scattering and RIXS experiments on two-leg ladder cuprate compounds, and our conclusions.

\section{Model}
The Hamiltonian of the Hubbard model defined on a two-leg ladder is
\begin{align}\label{eq:Hladder}
H&=\Big(-t_{x} \sum\limits_{\substack{\langle i,j\rangle\\ \sigma,\gamma}}
c^\dagger_{i,\gamma,\sigma}
c_{j,\gamma,\sigma} - t_y \sum\limits_{i,\sigma} 
c^{\dag}_{i,0,\sigma}c_{i,1,\sigma}\Big)+ \text{h.c.} \nonumber\\
&+ U\sum\limits_{i,\gamma} n_{i,\gamma,\uparrow}n_{i,\gamma,\downarrow},
\end{align}
where $c^\dagger_{i,\gamma,\sigma}$ ($c_{i,\gamma,\sigma}$) creates (annhilates) an electron 
at leg $\gamma=0,1$ on site $i=0,...,L/2-1$ and with spin $\sigma=\uparrow,\downarrow$.  
$L$ is the total number of sites, with $L/2$ sites on each leg, and 
$U$ is the strength of the Hubbard interaction.  
Following standard notation, $t_x$ and $t_y$ represent the nearest-neighbor hopping parameters in the $x$ (along the leg) and $y$
(along the rung) directions of the ladder. For simplicity, we denote the wavevector in the $y$ direction as $q_\text{rung} = 0,~\pi/a$ and the wavevector in the $x$ direction as $q$. 
For our DMRG calculations, we consider a ladder with open boundary conditions along the leg direction, while our RPA-FLEX calculations assume periodic boundary conditions along the leg direction and two sites in each rung are treated as two orbitals within each unit cell. In both cases, we adopt symmetric hopping integrals $t_x=t_y=t$. 
Throughout we take $t=1$ as our unit of energy and $a = 1$ as our unit of length. 

\section{Methods}
Many techniques ranging from exact diagonalization to DMRG~\cite{Noack94} to bosonization~\cite{re:Fabrizio1993,re:Balents1996,re:White2002} 
have been used to study the physics of the Hubbard two-leg ladder. However, to our knowledge, a 
comparison between the RPA treatment of a two-leg Hubbard ladder and an exact numerical approach like DMRG, has not been carried out. 

\subsection{RPA-FLEX}
In this section, we summarize the multi-orbital FLEX formalism used to compute the single particle and anomalous self-energies. Our notation follows that used 
in Refs.~\cite{takimoto2004strong, Yada2005, Zhang2009, Zhang2010}, which also provide a more detailed discussion of the formalism.   

The central quantities in the Eliashberg equations with FLEX interactions are the single particle $G_{l_1l_2}(k)$ and anomalous $F_{l_1l_2}(k)$ Green's functions, 
the single particle $\Sigma_{l_1l_2}(k)$ and anomalous $\Phi_{l_1l_2}(k)$ self-energies, and the 
particle-hole susceptibility $\chi_{l_1l_2l_3l_4}(q)$. Allowing for a nonzero 
anomalous self-energy is necessary to obtain meaningful 
results below the superconducting critical temperature $T_c$. This also simplifies the comparison with DMRG
calculations for the ground state. Above, 
$l_j$ are orbital-like indices ($l_j=1$ for leg 0 and $l_j=2$ for leg 1) and we have used the 4-vector 
notation with $k \equiv ({\bf k}, {\mathrm i}\omega_n)$ and $q \equiv ({\bf q},\mathrm{i}\omega_m)$ 
where $\omega_n=\frac{\pi}{\beta}(2n+1)$ and 
$\omega_m = \frac{\pi}{\beta}2m$ are used for fermion and boson Matsubara frequencies, respectively. 
For our two-leg ladder problem, we have a two-orbital unit cell (equivalent to a single rung of the ladder) 
and the Green's functions and self-energies are $2\times 2$ matrices in orbital space. 
For the particle-hole irreducible susceptibility, the four indices can be grouped as $A = (l_1l_2)$ and $B = (l_3l_4)$, 
such that $\chi_{A,B}(q)$ can be written as a $4\times 4$ 
matrix in orbital space with $(l_1l_2) = (11,22,12,21)$ for the rows and 
$(l_3l_4) = (11,22,12,21)$ for the columns 
\begin{equation}
\ubar{\chi}^p = \left( 
\begin{array}{cccc}
\chi^p_{11,11} & \chi^p_{11,22} & \chi^p_{11,12} & \chi^p_{11,21} \\
\chi^p_{22,11} & \chi^p_{22,22} & \chi^p_{22,12} & \chi^p_{22,21} \\
\chi^p_{12,11} & \chi^p_{12,22} & \chi^p_{12,12} & \chi^p_{12,21} \\
\chi^p_{21,11} & \chi^p_{21,22} & \chi^p_{21,12} & \chi^p_{21,21}
\end{array}
\right). 
\end{equation}
Here, we use the subscript $p = 0,s$ ($p = 0,c$) for the irreducible spin (charge) susceptibility, 
or $s$ ($c$) for the RPA spin (charge) susceptibility. 
The irreducible spin and charge susceptibilities are equal in the normal state 
but different in the superconducting state due to nonzero anomalous self-energies.
The dynamical spin and charge susceptibilities are respectively 
calculated from the RPA formula in a generalized matrix form as follows
\begin{eqnarray}\label{Eq:Chi}\nonumber
\ubar{\chi}^s(q)&=&\left[\mathbb{1} - \ubar{\chi}^{0,s}(q)\ubar{U}^s \right]^{-1}\ubar{\chi}^{0,s}(q),\\\nonumber
\ubar{\chi}^c(q)&=&\left[\mathbb{1} + \ubar{\chi}^{0,c}(q)\ubar{U}^c \right]^{-1}\ubar{\chi}^{0,c}(q),\\  
\end{eqnarray}
where $\mathbb{1}$ denotes a $4\times 4$ identity matrix, and $\ubar{U}^s$ and $\ubar{U}^c$ are the spin and charge 
interaction matrices. Note that this matrix-RPA form generates Feynman diagrams beyond the ring diagrams summed in the usual RPA formula~\cite{Altmeyer2016}.

Since the Hamiltonian Eq. (\ref{eq:Hladder}) only contains  
the on-site Hubbard interaction, the interaction matrices take a simple form 
\begin{equation}
\ubar{U}^s = \ubar{U}^c = \left( 
\begin{array}{cccc}
U & 0 & 0 & 0 \\
0 & U & 0 & 0 \\
0 & 0 & 0 & 0 \\
0 & 0 & 0 & 0
\end{array}
\right). 
\end{equation}

$\ubar{V}^n$ and $\ubar{V}^a$ define the effective FLEX interactions entering 
into the equations for the normal $\Sigma_{l_1 l_2}(k)$ and anomalous 
$\Phi_{l_1l_2}(k)$ self-energies, respectively. Due to the form of the interaction matrix used here, these have the simple form 
\begin{eqnarray}
\ubar{V}^n(q)&=&\frac{3 U^2}{2}\ubar{\chi}^s(q)+\frac{U^2}{2}\ubar{\chi}^c(q) - U^2 \ubar{\chi}^{0,G}(q) + U\mathbb{1},\\
\ubar{V}^a(q)&=&\frac{3 U^2}{2}\ubar{\chi}^s(q)-\frac{U^2}{2}\ubar{\chi}^c(q) - U^2 \ubar{\chi}^{0,F}(q) + U\mathbb{1}, 
\end{eqnarray}
where $\ubar{\chi}^{0,G} = (\ubar{\chi}^{0,s} + \ubar{\chi}^{0,c})/2$,  
$\ubar{\chi}^{0,F} = (\ubar{\chi}^{0,s} - \ubar{\chi}^{0,c})/2$ 
and each matrix is now defined in a $2\times 2$ subspace of the original two-orbital basis 
\begin{equation}
\ubar{\chi}^p = \left( 
\begin{array}{cc}
\chi^p_{11,11} & \chi^p_{11,22} \\
\chi^p_{22,11} & \chi^p_{22,22} 
\end{array}
\right). 
\end{equation}
The remaining susceptibilities do not enter into the formalism and do not need to be computed at this point. 
This means that the particle and the hole must be in the same orbital at the interaction vertex. 
For example, this happens in the particle-hole ring-diagram, where we do not have the interorbital Hubbard interaction in the Hamiltonian. 
In this case, interorbital propagation is still allowed because of the hopping 
along the rungs of the two-leg ladder, and 
the Green's functions are not diagonal in the orbital space.

Introducing the short-hand notation $\chi_{l,m}^{0,s}(q) \equiv
\chi_{ll,mm}^{0,s}(q)$, the irreducible spin (charge) susceptibilities are
given by  
\begin{eqnarray*}
\chi^{0,s}_{l,m}&=&-\frac{T}{N}\sum_k \left[
G_{lm}(k+q)G_{ml}(k) + F_{lm}(k+q)F^*_{ml}(k)\right], \\
\chi^{0,c}_{l,m}&=&-\frac{T}{N}\sum_k \left[
G_{lm}(k+q)G_{ml}(k) - F_{lm}(k+q)F^*_{ml}(k)\right],
\end{eqnarray*}
where $F^*$ denotes the complex conjugate of $F$. 
Since the FLEX interactions for our model Hamiltonian 
satisfy $V^{n(a)}_{ll^\prime,mm^\prime}(q) = V^{n(a)}_{l,m}\delta_{ll^\prime}\delta_{mm^\prime}$ 
the normal and anomalous self-energies can also be written in a compact form 
without any summation over the orbital index as
\begin{equation}\label{Eq:Sigma}
\Sigma_{lm}(k) = \frac{T}{N}\sum_{q}V^n_{l,m}(q)G_{lm}(k-q), 
\end{equation}
and 
\begin{equation}\label{Eq:Phi}
\Phi_{lm}(k) = \frac{T}{N}\sum_{q}V^a_{l,m}(q)F_{lm}(k-q). 
\end{equation}

Equations (\ref{Eq:Chi})-(\ref{Eq:Phi}) constitute the set of matrix FLEX equations, which we solve self-consistently together 
with Dyson's equation in the Nambu-orbital space. Since the momentum and frequency sums are in a convolution or cross-correlation form, 
we use fast Fourier transforms (FFT) to speed up the computation. 
 We use a $128\times 1$ $k$-grid and five times the bandwidth as the energy cutoff for the Matsubara frequencies.
During the self-consistent loop, 
we also adjust the chemical potential $\mu$ to keep the total electron filling $n$ fixed. The 
total density is computed from the electron Green's function as 
\begin{equation}
n = \frac{2T}{N}\sum_{l,{\bf k},n} G_{ll}({\bf k},{\mathrm i}\omega_n)e^{\mathrm{i}\omega_n 0^+}, 
\end{equation}
where $0^+$ denotes a positive infinitesimal number. 
Note that the Hartree-Fock contribution to the self-energy for our model is 
$\Sigma^{HF}_{11}(k) = Un^\sigma_{11}$ and $\Sigma^{HF}_{22}(k) = Un^\sigma_{22}$, which is independent of 
momentum and Matsubara frequency and independent of orbital index 
due to the degenerated orbitals. This contribution can therefore be absorbed into the 
chemical potential that is adjusted to fix the electron filling $n$.
A very low temperature $T=0.01t$ is used in RPA+FLEX calculations, 
except that at half filling $T=0.05t$ is used to avoid the magnetic instability due to the
tendency to antiferromagnetic order at low temperature.

\subsection{DMRG}
We employ the DMRG correction-vector method throughout this paper~\cite{re:Kuhner1999}. Within the correction vector approach, we use the Krylov 
decomposition~\cite{nocera2016spectral}
rather than the conjugate gradient. An application of the method to Heisenberg and Hubbard ladders at half-filling can be found in Ref.~\cite{nocera2016magnetic}, while Ref.~\cite{nocera2017signatures} presents a study of the pairing tendencies at finite hole-doping. 
In this work, a $L=48\times2$ ladder has been 
simulated, using $m=1000$ DMRG states with a truncation error kept below $10^{-5}$. 
The spectral broadening in the correction-vector approach 
was fixed at $\eta=0.08t$. The DMRG implementation used throughout 
this paper has been discussed in detail in~\cite{nocera2016magnetic}; 
technical details are in the Supplemental Material~\cite{re:supplemental}.

At each frequency $\omega$, we compute the dynamical spin structure factor of the two-leg ladder in real space
\begin{equation}\label{eq:Sijw}
S_{j,c}(\omega+i\eta)=\langle \Psi_{0}|S^{z}_{j} \frac{1}{\omega-H+E_{g}
+i\eta}S^{z}_{c}|\Psi_{0}\rangle,
\end{equation}
for all sites of the lattice, where $E_{g}$ is the energy of the 
ground state $|\Psi_{0}\rangle$ of the Hamiltonian $H$.
An analogous definition exists for the dynamical charge structure factor $N(\textbf{q},\omega)$, where the contribution from the static average densities is subtracted
\begin{equation}
\begin{aligned}\label{eq:Nijw}
N_{j,c}(\omega+i\eta)&=\langle \Psi_{0}|(n_{j}-\langle n_j\rangle)\frac{1}{\omega-H+E_{g}
+i\eta}\times\\
&\times (n_{c}-\langle n_c\rangle)|\Psi_{0}\rangle.
\end{aligned}
\end{equation}
Above, $j\equiv (j_x,j_\text{rung})$ corresponds to the two coordinates of the site on the ladder, where
$j_\text{rung}=0$ ($1$) for the lower (upper) leg of the ladder.
The center site is $c\equiv(L/4-1,0)$.
The above quantities are then Fourier transformed to momentum space giving two components 
(for brevity we report the formulas only for the dynamical spin structure factor)
\begin{equation}\label{eq:Sladw}
\begin{aligned}
S((q &,q_\text{rung}=0),\omega)=\sqrt{\frac{2}{L/2+1}}\sum_{j_x=0}^{L/2-1}\sin((j_x+1) q)\times \\
&\times \big[S_{(j_x,0),c}(\omega+i\eta)+S_{(j_x,1),c}(\omega+i\eta)\big],\\
S((q &,q_\text{rung}=\pi),\omega)=\sqrt{\frac{2}{L/2+1}}\sum_{j_x=0}^{L/2-1}\sin((j_x+1) q)\times \\
&\times \big[S_{(j_x,0),c}(\omega+i\eta)-S_{(j_x,1),c}(\omega+i\eta)\big],
\end{aligned}
\end{equation}
where the quasi-momenta $q=\frac{\pi n}{L/2+1}$ with $n=1,..,L/2$ are appropriate for 
open boundary conditions on each leg. 

\section{Results}

\subsection{Ground state pairing properties}
We begin by studying the ground state pairing properties obtained with DMRG and RPA+FLEX (the latter at low but finite temperature) approaches.

\begin{figure}[h]
\includegraphics[width=8.5cm]{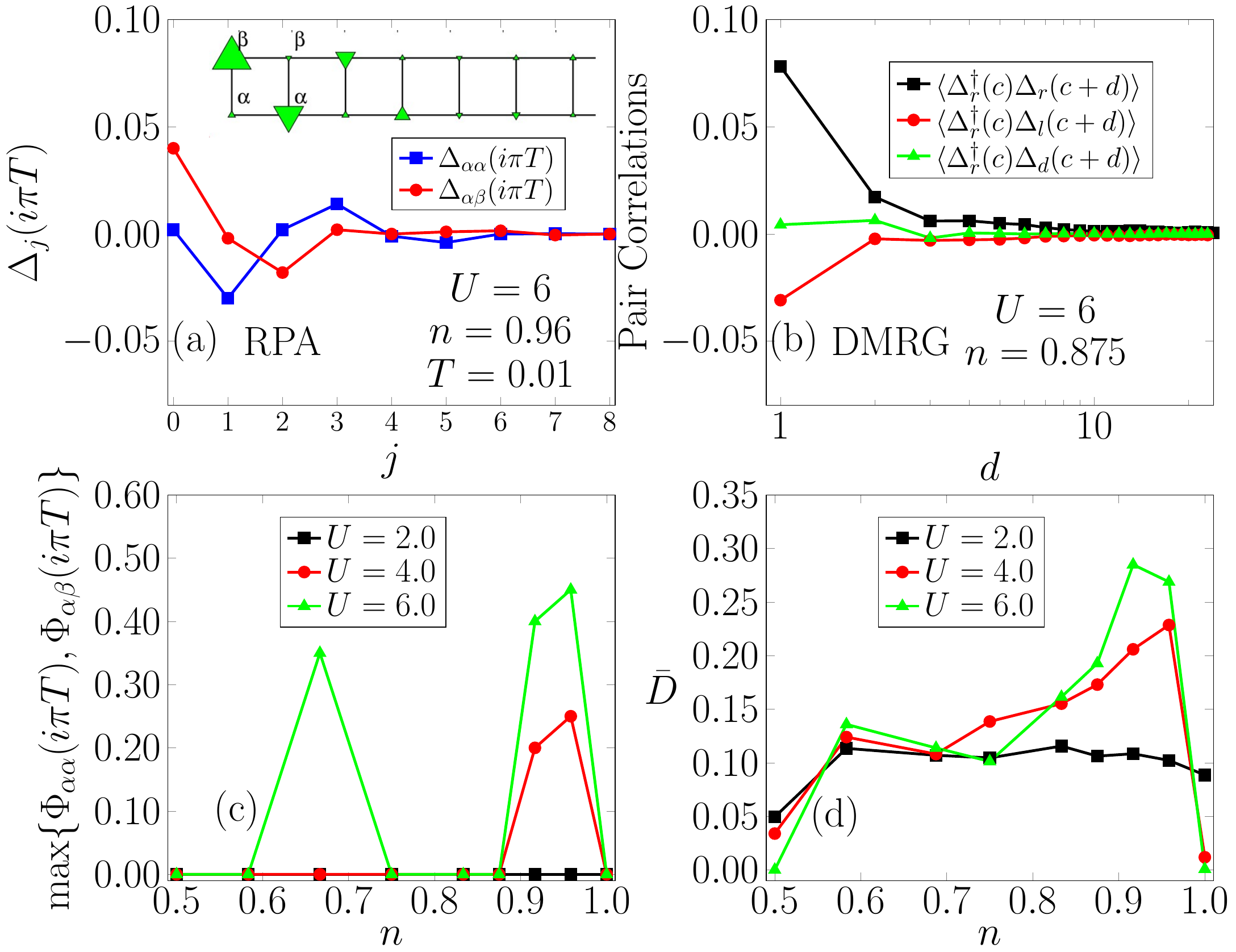}
\caption{(a) Superconducting gap function (in units of $t$) computed with RPA+FLEX 
as a function of space and leg index 
($\alpha$ corresponds to lower leg, while $\beta$ to upper leg). Here, $U/t=6.0$, 
electron filling $n=0.96$. 
The inset in (a) is a pictorial representation of the pairing gap 
at first few sites of the two-leg ladder, with one of the electron
fixed at site-$0$ of the lower leg. The upward triangle means a positive gap and the downward triangle means a negative gap and
the size of the triangle is proportional to the gap magnitude.
(b) Rung-rung, rung-leg, 
and rung-diagonal pair singlet correlation functions computed with DMRG as a function 
of the distance from the center of the ladder. Here, $U/t=6.0$, $n=0.875$.
(c) Maximal anomalous self-energy (in units of $t$) in the first Brillouin 
zone computed in RPA+FLEX as a function of electron filling and different values of $U$, as indicated.
(d) Pairing strength computed with DMRG as a function of electron filling, 
for different values of $U$, as indicated. The pairing strength is computed from 
the rung-rung pair singlet correlation functions as $\bar{D}=\sum_{j=6}^{j=12}P(j)/P(1)$, 
where $P(j)=\langle\Delta^{\dag}_{r}(c)\Delta_{r}(c+j)\rangle$.
The persistent background at $U/t=2$ over a wide range of doping originates 
in short distance correlations even in the non-interacting limit.} \label{fig7}
\end{figure}

Figure~\ref{fig7}a shows the RPA+FLEX superconducting gap as a function of space index $j$ and leg index ($\alpha$ for leg 0 and $\beta$ for leg 1),
indicating the $d$-wave-like character of the superconducting ground state, which is characterized by a non-zero order parameter
at sufficiently low temperatures and a gap sign change between site $(j_x,j_\text{rung})=(1,0)$ and $(0,1)$.
As opposed to the RPA approach, that works in the grand canonical ensemble, 
our finite-size DMRG simulations are performed at fixed number of electrons present in the system, thus
one cannot have a non-zero superconducting order parameter $\langle\Delta_{r}(i)\rangle$, 
where 
\begin{equation}\label{deltarung}
\Delta_{r}(i) = \frac{1}{\sqrt{2}}\Big(c_{i,0,\uparrow}c_{i,1,\downarrow}-c_{i,0,\downarrow}c_{i,1,\uparrow}\Big)
\end{equation}
for local singlet operators on a rung of the ladder. 
However, DMRG calculations have shown that in the weakly hole-doped regime the doped Hubbard ladder exhibits 
dominating superconducting tendencies: rung-siglet superconducting correlations have the \emph{slowest} 
power-law decay as a function of distance~\cite{PhysRevB.92.195139}. This is the typical behavior of quasi-one dimensional systems, 
and one assumes that the system is quasi-long-range ordered. 
DMRG computations have also shown that superconducting quasi-order has $d$-wave-like character. 
We report the results showing this behavior in Fig.~\ref{fig7}(b), which shows the pair-pair singlet 
correlations as a function of the distance $d$ along the leg of the ladder, fixing the Hubbard repulsion to strong coupling $U/t=6$ and the electron filling to $n=0.875$. 
We first fix the creation of a singlet pair of electrons on a rung at the center of the ladder 
(see the definition of the \emph{destruction} operator in Eq.~(\ref{deltarung})). 
We then consider three different possibilities for the pair-pair correlations by destroying the pair (1) along a rung [Eq.(\ref{deltarung})], (2) 
along diagonal, and (3) along a leg at a distance $d$ from the center. The operators destroying singlet pairs along the last two directions
at a position $i$ on the ladder are defined as follows:  
\begin{equation}
\begin{aligned}
\Delta_{d}(i) &= \frac{1}{\sqrt{2}}\Big(c_{i,0,\uparrow}c_{i+1,1,\downarrow}-c_{i,0,\downarrow}c_{i+1,1,\uparrow}\Big),\\
\Delta_{l}(i) &= \frac{1}{\sqrt{2}}\Big(c_{i,0,\uparrow}c_{i+1,0,\downarrow}-c_{i,0,\downarrow}c_{i+1,0,\uparrow}\Big).
\end{aligned}
\end{equation}

Pair-pair correlations are $d$-wave-like, showing a change of sign going from 
the rung-rung to the rung-leg directions. This result agrees with the $d$-wave character of the superconducting ground state found in RPA+FLEX.
Within the RPA approach, the superconducting pairing strength can be quantified by evaluating the maximum of the anomalous 
self-energy (see Fig.~\ref{fig7}c). For low hole-doping ($\leqslant 10\%$), pairing tendencies increase when the Hubbard  
repulsion strength $U/t$ is increased above intermediate values, $U/t\simeq 3$. 

Moreover, notice the occurence of a non-zero \emph{peak} 
in the maximal anomalous self-energy for electron filling $n=0.666$ and strong Hubbard repulsion $U/t=6$. 
Unlike the pairing state in low hole-doping cases, where $\langle\Delta_l\rangle$ and 
$\langle\Delta_r\rangle$ have opposite sign but similar magnitude from the RPA+FLEX calculation, 
for $n=0.666$ and $U=6$ one has $|\langle\Delta_l\rangle| \ll |\langle\Delta_r\rangle|$, i.e., 
the pairing along the rungs dominates. 
The result at this filling $n$ is reproducible with larger $k$-grid, higher frequency cut-off, 
and stronger $U$ (no pairing for $U/t \geq 10$, however) 
in the RPA+FLEX calculations, but the pairing is quite sensitive to even a small deviation to the filling $n$, 
which does not coincide with quarter filling $n=0.5$. (The
van Hove singularity gives diverging density of states at the Fermi level of the \emph{noninteracting} bands at quarter filling.)
The physical nature of this peak in 
the maximal anomalous self-energy within our FLEX approximation is under investigation. 

Figure~\ref{fig7}(d) computes the 
pairing correlation strength with DMRG, which we estimate by evaluating the quantity $\bar{D}=\sum_{i=6}^{12}P(i)/P(1)$. (Note that $6$ and $12$ 
are arbitrary lower and upper bounds in the sum. The results are qualitatively similar if we modify these bounds; choosing $6$, 
as opposed to, {\it e.g.} 1, reduces artificial short-distance effects while $12$, as opposed to, {\it e.g.} 24, reduces edge effects.) 

Similar to RPA, DMRG results also show that pairing intensities are robust up to an electron doping which is close to $n\simeq0.6$. 
Except for the \emph{anomalous} peak in the RPA self-energy, 
we observe overall a good qualitative agreement between the pairing strength evolution 
with doping found in DMRG and the maximum of anomalous self-energy computed within the RPA+FLEX approach. 
In particular, pairing tendencies for small hole-doping intensify as one increases the Hubbard $U$ interaction from weak to strong coupling. 
In fact, low-energy charge fluctuations are suppressed while spin fluctuations become more robust for an increasing  Hubbard $U$. 
In this regime, hole pairing along the rungs of the ladder dominates~\cite{re:Dagotto1992}.

\subsection{Spin and charge excitations at weak coupling}

\begin{figure}[h]
\includegraphics[width=8.5cm]{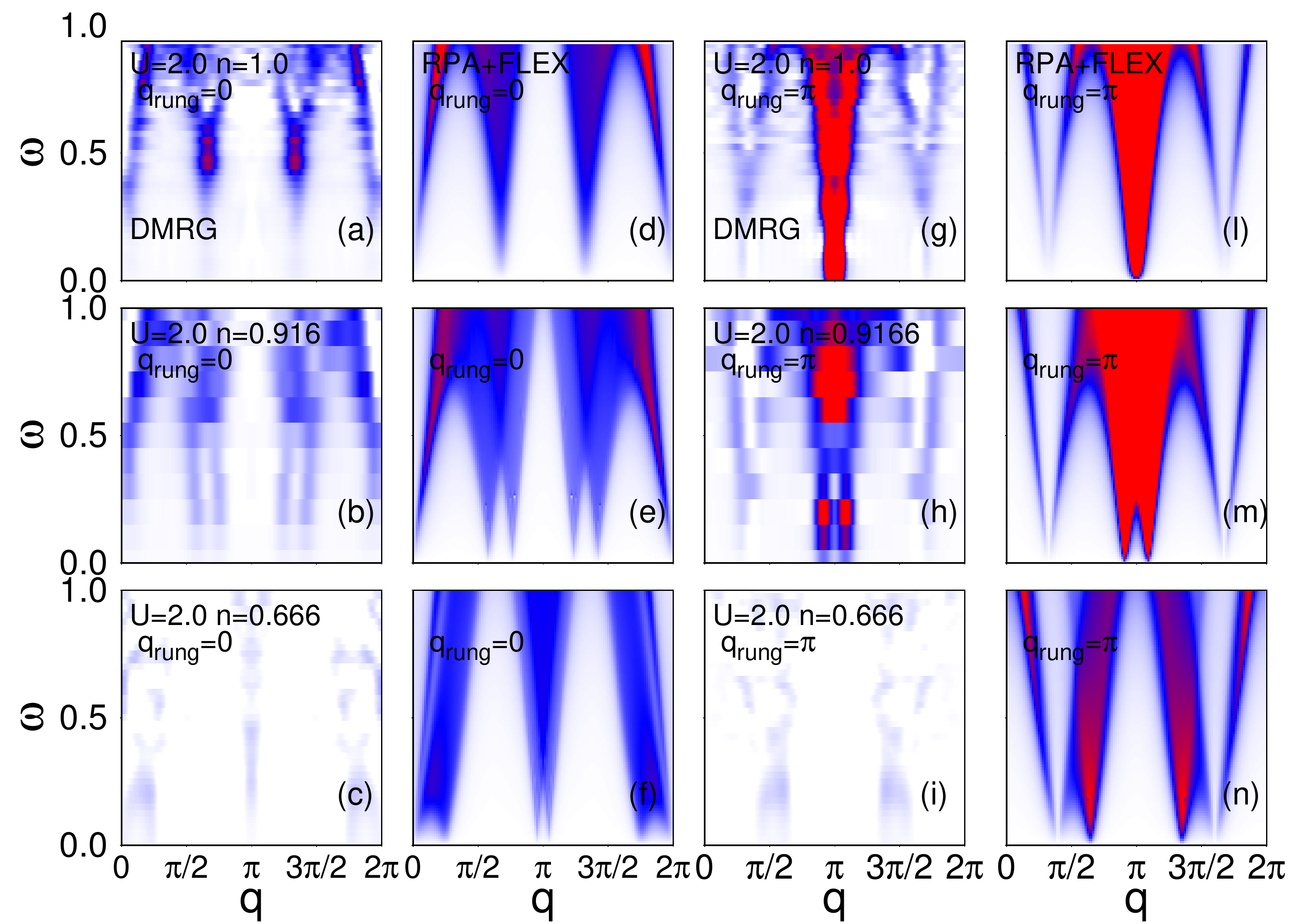}
\caption{Magnetic excitation spectrum $S({\bf q},\omega)$ for a $L=48\times 2$ 
ladder from DMRG (panels (a-c) for $q_\text{rung}=0$, panels (g-i) for $q_\text{rung}=\pi$) and RPA+FLEX 
(panels (d-f) for $q_\text{rung}=0$, and (l-n) for $q_\text{rung}=\pi$). 
$U/t=2.0$, as indicated. The electron doping $n=N/L$ is shown in each panel.
DMRG used $m=1000$ states and $\eta=0.08$.
RPA also used $\eta=0.08$. 
RPA used Pad{\'e} analytic continuation to obtain the complex function $S({\bf q},\omega+i\eta)$. 
In RPA, $q_\text{rung}=0~(\pi)$ component is obtained from $\chi_{+(-)}^s=\chi_{+(-)}^{0,s}/(1-U\chi_{+(-)}^{0,s})$, where $\chi_{+(-)}^{0,s}
=\chi_{1,1}^{0,s}+(-) \chi_{1,2}^{0,s}$. Here $+(-)$ denotes the $q_\text{rung}=0(\pi)$ component.} \label{fig1}
\end{figure}

\begin{figure}[h]
\includegraphics[width=8.5cm]{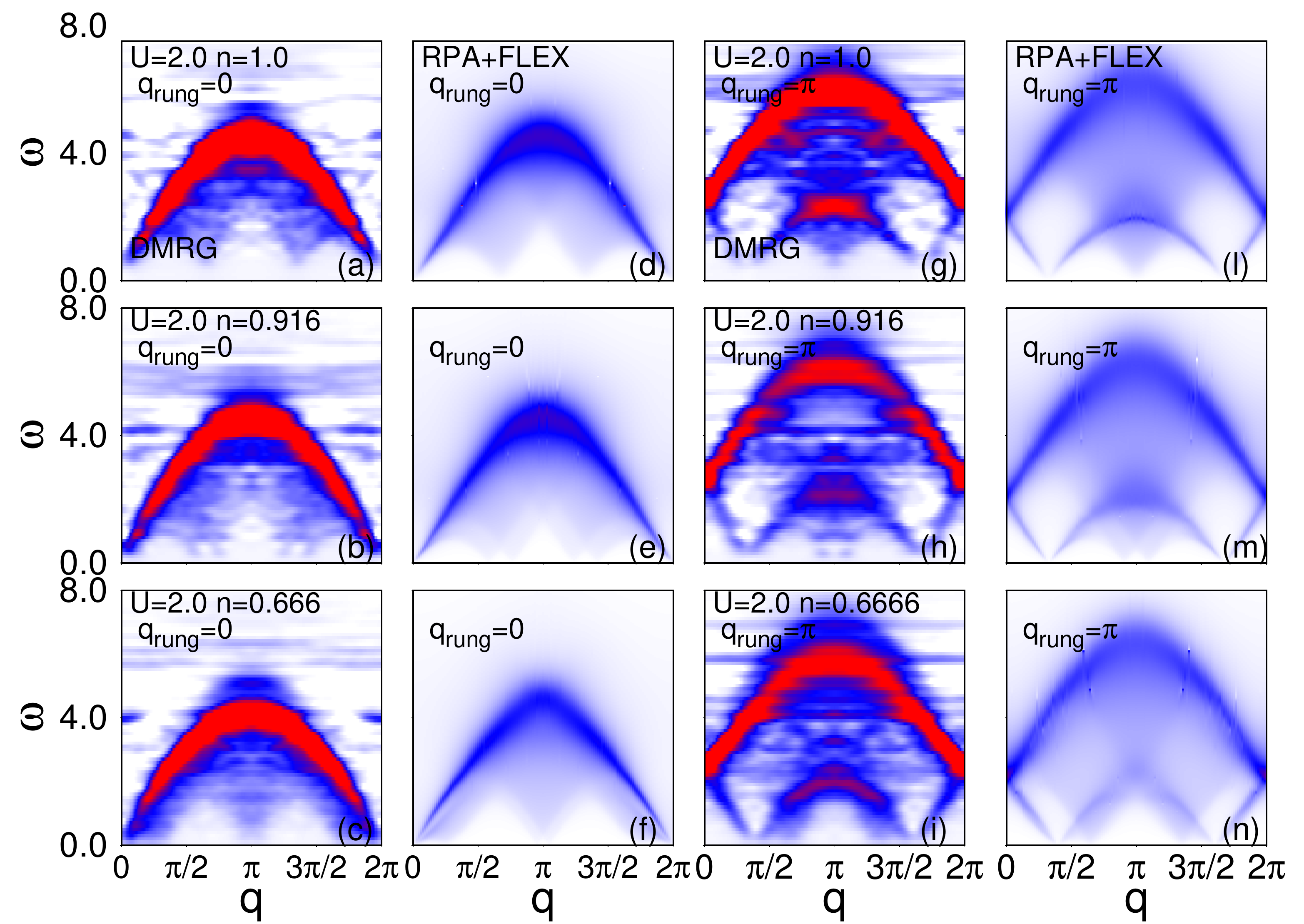}
\caption{Charge excitation spectrum $N({\bf q},\omega)$ for a $L=48\times 2$ 
ladder from DMRG (panels (a-c) for $q_\text{rung}=0$, panels (g-i) for $q_\text{rung}=\pi$) 
and RPA+FLEX (panels (d-f) for $q_\text{rung}=0$, and (l-n) for $q_\text{rung}=\pi$). 
$U/t=2.0$, as indicated. The electron doping $n=N/L$ is shown in each panel.
DMRG used $m=1000$ states and $\eta=0.08$. 
RPA used also $\eta=0.08$ in the Pad{\'e} analytic continuation.} \label{fig2}
\end{figure}

Figures~\ref{fig1} and \ref{fig2} display the spin and charge dynamical structure factors, respectively, for our two-leg Hubbard ladder in the
weak Hubbard $U$ regime ($U/t=2$) for three different values of the electron filling: half-filled $n=1.0$, corresponding to the undoped regime;
$n=0.9166$, corresponding to the weakly hole-doped regime ($\simeq 8\%$); finally $n=0.666$, corresponding to a heavily hole-doped 
regime ($\simeq 33\%$). In each figure, spectra computed with DMRG appear in panels (a-c) (with the response along the direction $(q,0)$ in the 
Brillouin zone reported) and in panels (g-i) (with the momentum along the 
direction $(q,\pi)$). Analogously, the panels (d-f) and (l-n) report 
the spectra along the same momentum directions computed with RPA+FLEX approximation.

At weak Hubbard repulsion ($U/t=2$), RPA+FLEX calculations well reproduce the magnetic excitation spectra computed with DMRG. 
In the $q_\text{rung}=\pi$ component in the undoped regime (panels (g) and (l)), one can observe the typical one-magnon V-shape-like dispersion 
around $(\pi,\pi)$, where the majority of the spectral weight is located.
Notice that, even though the spectral weight is already concentrated at low energy for $U/t=2$~\cite{nocera2016magnetic}, 
the side branches corresponding to weakly interacting electron-hole excitations across the 
``Fermi surface'' (which become gapless at scattering momenta $q\simeq\pi/3$ and $q\simeq2\pi-\pi/3$) are correctly captured by RPA.
In the $q_\text{rung}=0$ component, the dispersion and spectral weight of magnetic excitations,   
which correspond to intraband electron-hole excitations in the $U/t=0$ case, are also correctly reproduced. 
The DMRG results, however, seem to indicate that a pseudogap for momentum transfers around $q=(\pi,0)$ 
is already forming (panels (a) and (d) of Fig.~\ref{fig1}). 

In the weakly doped regime, incommensurate peaks at positions
proportional to the electronic density develop around $(\pi,\pi)$ (see also Ref.~\cite{nocera2017signatures}). 
In this frequency-momentum region, also notice
the difference in spectral weight distribution between DMRG in panel (h) and RPA in panel (m): RPA shows that the magnetic spectral intensity is even more substantial at very low energy, while the DMRG results show a maximum around $\omega\simeq0.6t$.  
A similar behavior is observed for the
gapless magnetic excitation branches at $q \simeq (\pi\pm\pi/3,0)$ (see panels (b) and (e)).
These follow closely the dispersion of intraband electron-hole excitations in the $U/t=0$ case, as observed in the undoped regime.

In the overdoped regime, $n=0.666$ (bottom row of panels in Fig.~\ref{fig1}), the RPA+FLEX approximation correctly captures the dispersion of magnetic excitations, 
which behave as weakly interacting electron-hole excitations. Notice the difference in spectral weight intensity between DMRG and RPA results:
the spectra along both directions in the Brillouin zone are plotted using the same color intensity, and this makes the DMRG result appear very weak. 
In particular, RPA overestimates the spectral weight of the magnetic excitations, as was the case for small doping. 

We now discuss the charge excitations reported in Fig.~\ref{fig2}: for all the dopings investigated $N(\textbf{q},\omega)$ computed with DMRG are well captured by RPA. 
In particular, RPA describes well the gapless excitations and the concentration of spectral weight at high energy in 
both $q_\text{rung}=0$ and $q_\text{rung}=\pi$ components. However, as opposed to the case of 
the magnetic spectra, DMRG predicts a more substantial spectral weight than RPA. 

The spectral features shown by DMRG and RPA can be easily understood in terms
of non-interacting electron-hole excitations across the Fermi surface of the ladder. 
Notice that, for $U/t=0$ and in the undoped regime $n=1.0$,
both anti-bonding (higher energy) and bonding (lower energy) bands are partially filled by electrons
with filling $n_1=1/3$ ($k_\text{F}=\pi/3$ measured from $k=0$) and $n_2=2/3$ ($k_\text{F}=2\pi/3$ 
measured from $k=0$), respectively. The charge response along the 
direction $(q,\pi)$ corresponds to excitations \emph{across} bonding and antibonding bands. 
These describe the prominent excitation \emph{arc} starting from $q=0$ and $\omega\simeq2t$, reaching a 
maximum for $q=\pi$ and $\omega\simeq6t$, where electrons from the bottom of the bonding band are excited to the top of the anti-bonding band (see panels (g) and (l)). 

The low energy part of the spectrum has a mushroom-like shape, and describe electron-hole excitations within the energy interval $2t_y$ giving the energy separation between bonding and anti-bonding bands. Notice that electrons in the partially
filled anti-bonding band can be excited to states in the bonding band for small energy and large momentum transfers as well.
One can observe finally the presence of gapless excitations for momenta $(\pi,\pi)$, $(k^*,\pi)$, and $(2\pi-k^*,\pi)$ with $k^*\simeq\pi/3$.
These correspond to the minimum and maximum momentum transfer allowed at zero energy 
for electron-hole excitations, respectively.
The charge response along the direction $(q,0)$, corresponds at $U/t=0$ to electron-hole excitations \emph{within} the bands of the ladder, 
which are both partially filled as stated above. 

\begin{figure}[h]
\includegraphics[width=8.5cm]{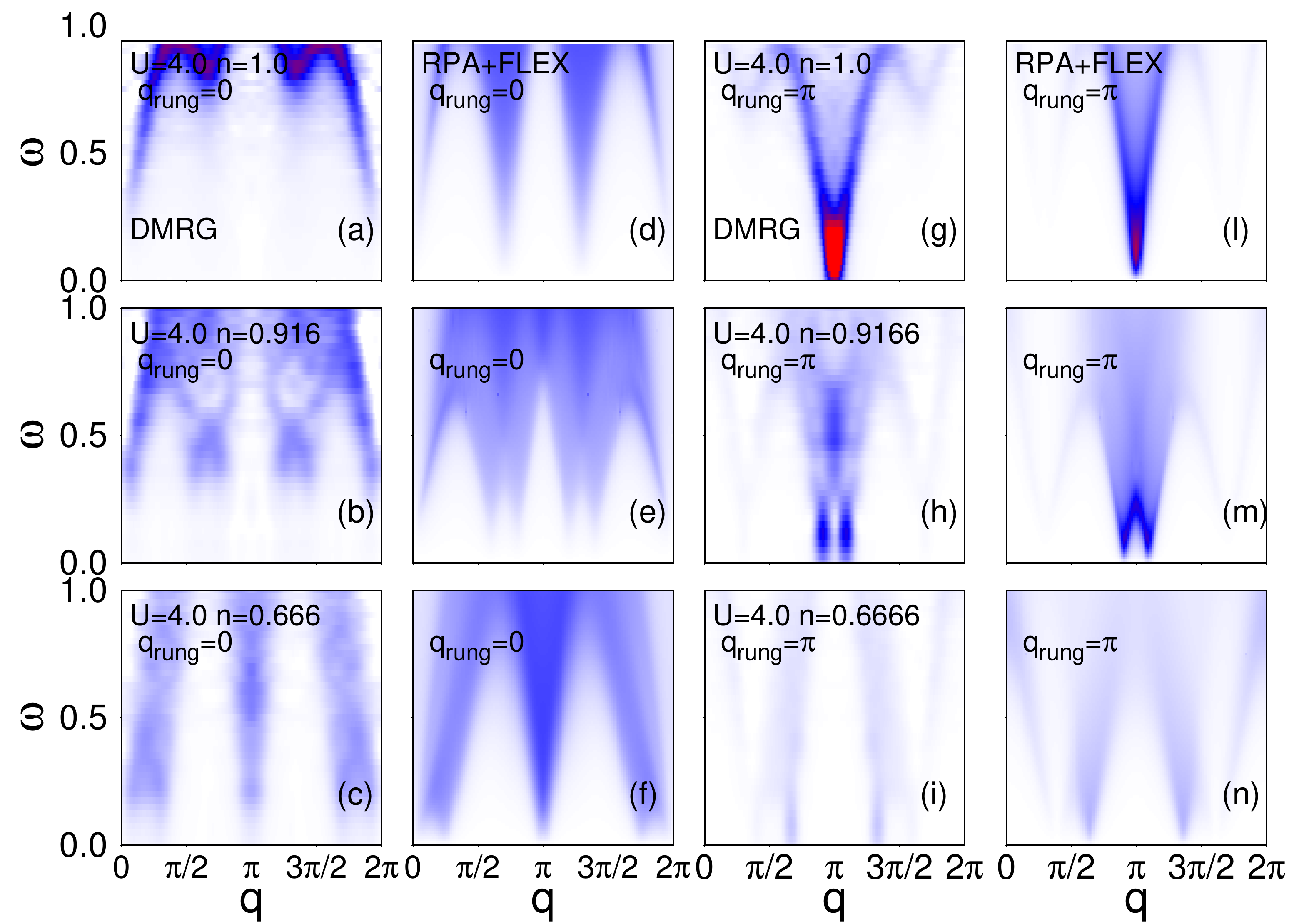}
\caption{Magnetic excitation spectrum $S({\bf q},\omega)$ for a $L=48\times 2$ 
ladder from DMRG (panels (a-c) for $q_\text{rung}=0$, panels (g-i) for $q_\text{rung}=\pi$) and RPA+FLEX 
(panels (d-f) for $q_\text{rung}=0$, and (l-n) for $q_\text{rung}=\pi$). 
$U/t=4.0$, as indicated. The electron doping $n=N/L$ is shown in each panel.
DMRG used $m=1000$ states and $\eta=0.08$. RPA used also $\eta=0.08$ 
in the Pad{\'e} analytic continuation.} \label{fig3}
\end{figure} 
 
At finite hole-dopings, much of the observations given above can be repeated. Notice, however, the appearance of incommensurate peaks around 
$(\pi,\pi)$ (see panels (g-n)), which also change position as a function of electron filling, similarly 
to the case in the magnetic excitation spectra. At the same time, both DMRG and RPA confirm that Fermi-surface effects 
give incommensurate peaks around 
$(q=2k_\text{F},0)=(2\pi/3,0)$ and $(4\pi/3,0)$ (see panels (a-f)).

\subsection{Spin and charge excitations at intermediate coupling}

In the regime of intermediate Hubbard $U$ ($U/t=4$), the main features of the magnetic excitation 
spectra are also well captured by RPA for all the dopings investigated, as shown in Fig.~\ref{fig3}.

\begin{figure}[h]
\includegraphics[width=8.5cm]{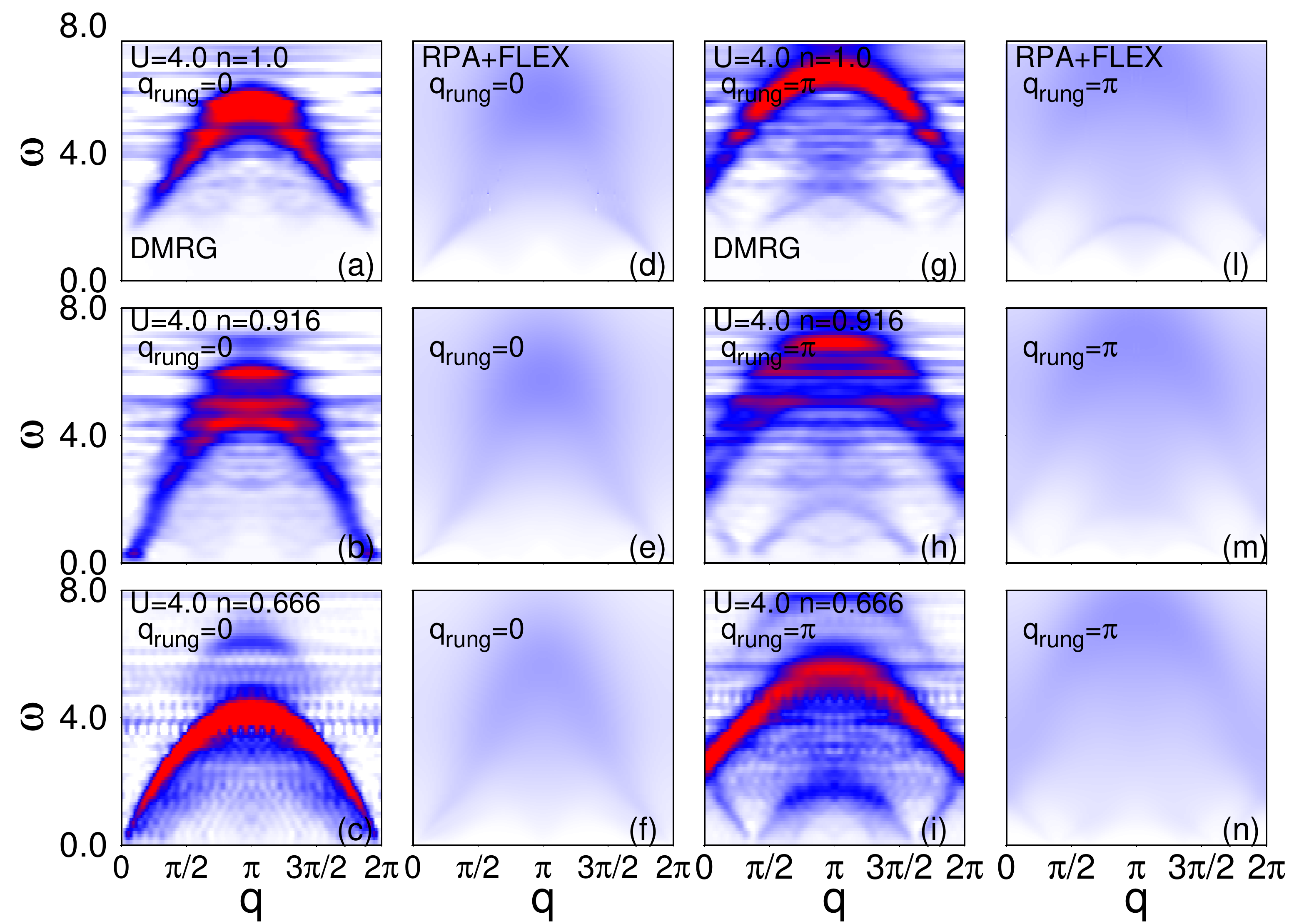}
\caption{Charge excitation spectrum $N({\bf q},\omega)$ for a $L=48\times 2$ 
ladder from DMRG (panels (a-c) for $q_\text{rung}=0$, panels (g-i) for $q_\text{rung}=\pi$) and RPA+FLEX 
(panels (d-f) for $q_\text{rung}=0$, and (l-n) for $q_\text{rung}=\pi$). 
$U/t=4.0$, as indicated. The electron doping $n=N/L$ is shown in each panel.
DMRG used $m=1000$ states and $\eta=0.08$.
RPA used also $\eta=0.08$ in the Pad{\'e} analytic continuation.} \label{fig4}
\end{figure}
In the undoped regime (panels (g) and (l)), we again observe a V-shape-like dispersion band around $(\pi,\pi)$, 
where the majority of the spectral weight is concentrated. 
However, while side branches corresponding to weakly interacting electron-hole excitations across the ``Fermi surface''
appear still gapless or weakly gapped at scattering momenta $q\simeq\pi/3$ and $q\simeq2\pi-\pi/3$ in RPA+FLEX, 
these are gapped in the DMRG spectra.
We can explain this behavior by observing that larger Hubbard $U$ couplings start 
to affect first large momentum transfers in electron-hole quasi-particle excitations.
Analogously, in the $q_\text{rung}=0$ component, the dispersion of the magnetic 
excitation branches at $q\simeq \pi\pm\pi/3$ appear gapped in the DMRG spectral 
while they remain gapless in RPA (panel (a) and (d) of Fig.~\ref{fig3}). 
Both DMRG and RPA give a gapped spectrum at $q=0$ in the $q_\text{rung}=0$ component, however.
In the weakly doped regime, discrepancies between DMRG and RPA magnetic spectra greatly reduce, 
both in the $q_\text{rung}=\pi$ and $q_\text{rung}=0$ components.
Finally, an excellent agreement between DMRG and RPA results is observed in the overdoped regime, $n=0.666$ (panels (c)-(f)-(i)-(n)). 
As in the weak Hubbard $U$ regime, we notice a discrepancy in the spectral weight of the magnetic excitations between DMRG and RPA. 
Specifically, for all the dopings investigated at intermediate $U$, 
DMRG reports a slightly higher magnetic spectral weight of the magnetic excitations in the $q_\text{rung}=0$ component.
Instead, in the $q_\text{rung}=\pi$ component, RPA reports a magnetic spectral weight in very good agreement with DMRG spectra. 

Next, we consider the charge excitations spectra in Figure~\ref{fig4}. In the undoped case, 
we can observe in the DMRG results (panel (a) and (g)) that a more substantial Mott charge gap is present in 
the system in both the $q_\text{rung}=0,\pi$ components. 
However, the RPA+FLEX approach misses this information, where we can only observe an incoherent band of excitations above some low energy excitations which are still gapless.
The picture that emerges from the DMRG-RPA comparison improves slowly with doping. In the large doping regime, one can see that the RPA 
approach begins to capture the low energy behavior of the DMRG spectra correctly. 
The high energy bands deviate less significantly from the DMRG results. We have verified that only at larger hole dopings ($\simeq 50\%$) 
we start to see good qualitative agreement between DMRG and RPA results. Overall, the RPA+FLEX significantly underestimates the dynamical charge response comparing to DMRG. We stress that the magnitude of $N(\mathbf{q},
\omega)$ is much smaller than $S(\mathbf{q},\omega)$ from both DMRG and RPA+FLEX calculations, 
and this indicates that pairing is dominated by the spin-fluctuations.

\begin{figure}[h]
\includegraphics[width=8.5cm]{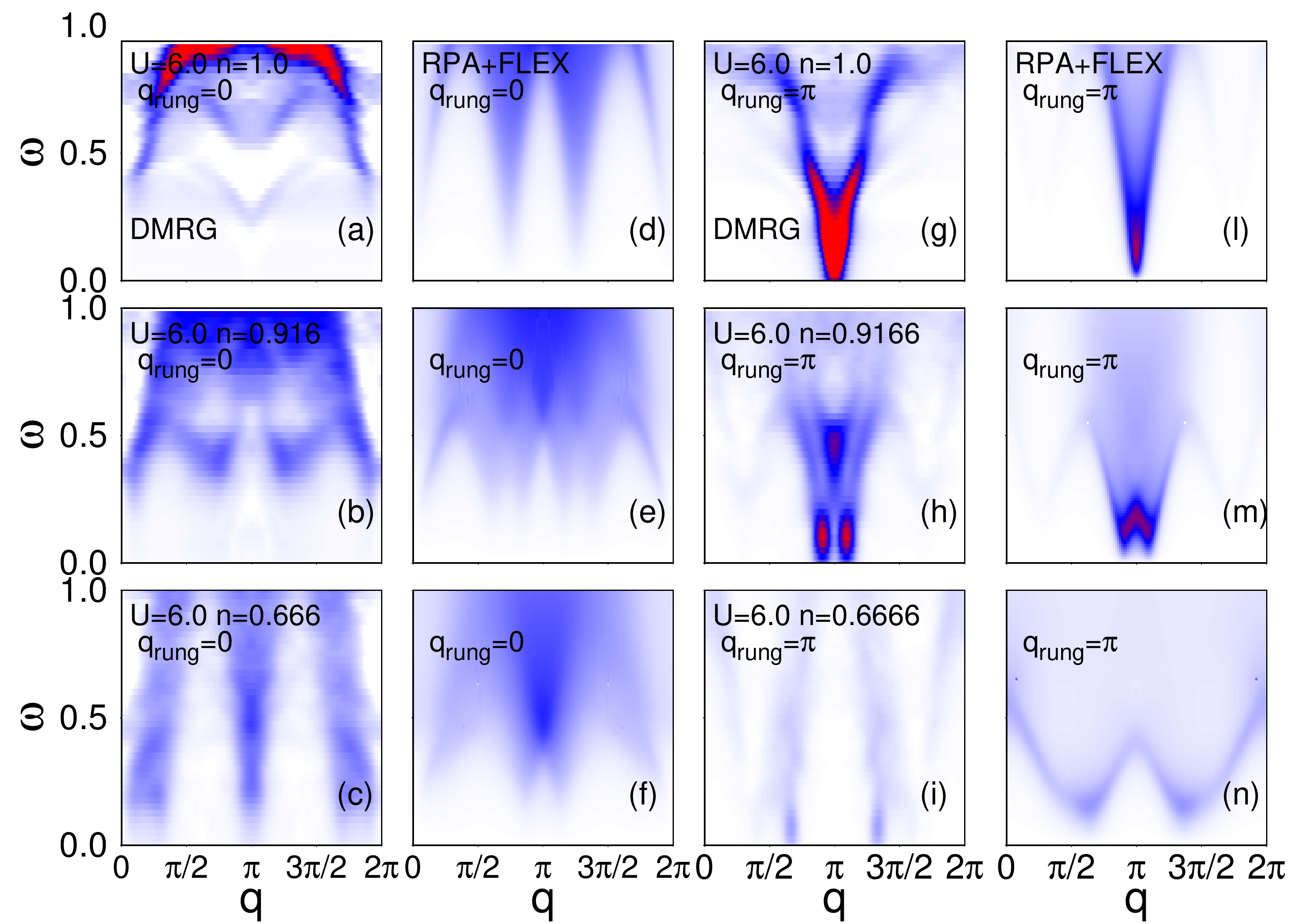}
\caption{Magnetic excitation spectrum $S({\bf q},\omega)$ for a $L=48\times 2$ ladder
from DMRG (panels (a-c) for $q_\text{rung}=0$, panels (g-i) for $q_\text{rung}=\pi$) and RPA+FLEX 
(panels (d-f) for $q_\text{rung}=0$, and (l-n) for $q_\text{rung}=\pi$). 
In this figure $U/t=6.0$, as indicated. The electron doping $n=N/L$ is shown in each panel.
DMRG used $m=1000$ states and $\eta=0.08$. 
RPA used also $\eta=0.08$ in the Pad{\'e} analytic continuation.} \label{fig5}
\end{figure}

\subsection{Spin and charge excitations at strong coupling}
We finally consider the strong Hubbard $U$ limit ($U/t=6$). 
In this case, both magnetic and charge excitation spectra 
computed with RPA present qualitative differences from the spectra computed with DMRG, as expected. 

In the $q_\text{rung}=\pi$ component in the undoped regime (panels (g) and (l) in Figure~\ref{fig5}), one can 
again observe a V-shape-like dispersion around $(\pi,\pi)$ in both DMRG and RPA+FLEX magnetic excitation spectra.
However, we notice that the spectral weight distribution is different, while 
at intermediate, up to high energies, the dispersion of the magnetic excitations are completely different in the two approaches. 
At finite doping, the agreement between DMRG and RPA+FLEX does not improve significantly: in the weakly doped regime, 
both the $q_\text{rung}=0$ and $q_\text{rung}=\pi$ spectra span along the same interval
of energies. However, the dispersion of low energy excitations is qualitatively different in the entire Brillouin zone.
In the large doping regime, the situation for the $q_\text{rung}=\pi$ component of the spectrum is very different:
RPA+FLEX spectrum is gapped in both $q_\text{rung}=0,\pi$ components, while DMRG shows gapless excitations. 
Last, we only begin to see qualitative similarities between the two approaches for the $q_\text{rung}=0$ spectra at large doping.
We also mention a difference between DMRG and RPA+FLEX approached at low $T$: 
while pairing fluctuations are included, there is no finite pairing order in
DMRG because a finte lattice size is used; for RPA+FLEX, the anomalous 
self-energies are not zero and may affect the $S(\mathbf{q},\omega)$ shown in panel (e)-(m)-(f)-(n).
\begin{figure}[h]
\includegraphics[width=8.5cm]{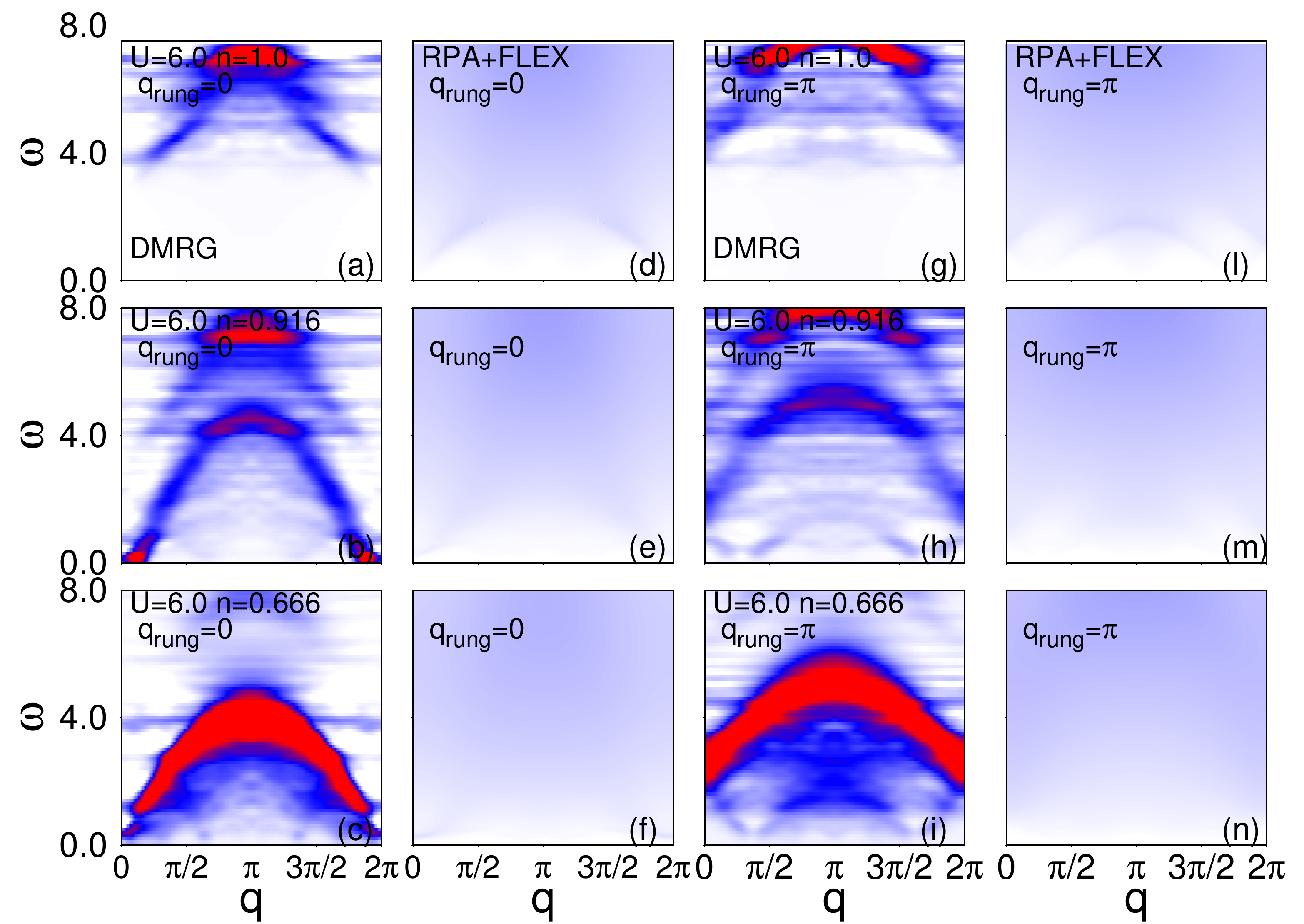}
\caption{Charge excitation spectrum $N({\bf q},\omega)$ for a $L=48\times 2$ ladder 
from DMRG (panels (a-c) for $q_\text{rung}=0$, panels (g-i) for $q_\text{rung}=\pi$) and RPA+FLEX 
(panels (d-f) for $q_\text{rung}=0$, and (l-n) for $q_\text{rung}=\pi$). 
$U/t=6.0$, as indicated. The electron doping $n=N/L$ is shown in each panel.
DMRG used $m=1000$ states and $\eta=0.08$. 
RPA used also $\eta=0.08$, in the Pad{\'e} analytic continuation.} \label{fig6}
\end{figure}
Finally, we consider the charge excitations spectra in Fig.~\ref{fig6}. In the RPA approach, 
the $N(\textbf{q},\omega)$ spectrum looks completely incoherent and featureless. 
Instead, DMRG results show that the spectra are rich, with both high energy bands above the Mott gap, and dispersive gapless excitations. 
For large $U$, the RPA+FLEX approximation fails to give an accurate result for the dynamical charge response, which is an order of magnitude smaller than the dynamical spin response according to DMRG.

\section{Discussion and Conclusions}
\begin{figure}[h]
\includegraphics[width=8.5cm]{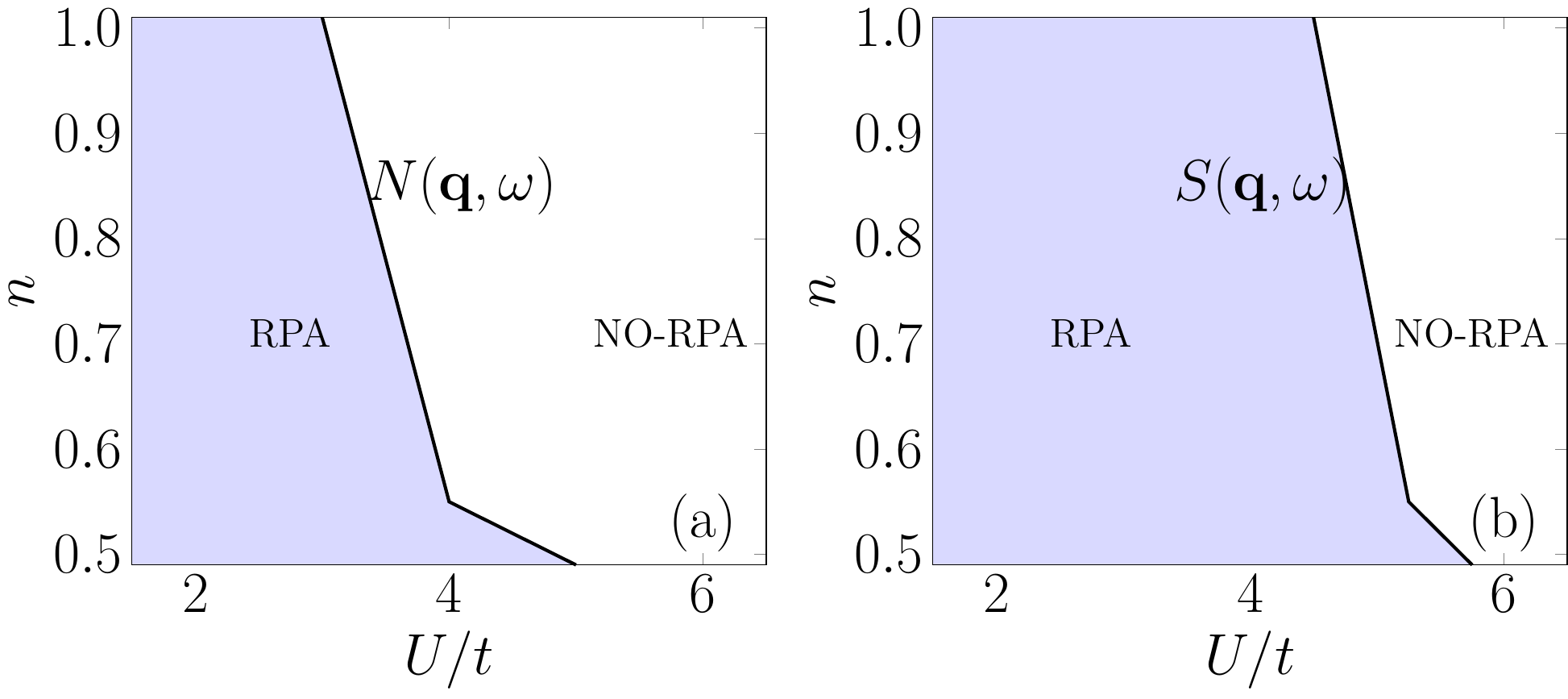}
\caption{(a) Sketch of the range of qualitative agreement between RPA+FLEX 
approximation for $N(\textbf{q},\omega)$, 
when compared with  numerically exact DMRG results. (b) Same as in panel (a) but referred to $S(\textbf{q},\omega)$. 
Notice that RPA-DMRG qualitative agreement range is \emph{larger} for $S(\textbf{q},\omega)$, in the range of parameter 
investigated in this work.} \label{fig8}
\end{figure}

Figure~\ref{fig8} summarizes our results in a diagram of the region of
$n$-$U/t$ parameter space where we find qualitative agreement between RPA+FLEX approximation and 
numerically exact DMRG results. From the analysis, it has emerged that the RPA+FLEX
approach works better for magnetic than charge excitations. Nevertheless, we
found that spin excitations are affected: the magnetic excitations
became \emph{more} gapped by increasing values of the Hubbard $U$, and only in
the large $U$ regime became qualitatively different from the spectrum produced by
weakly interacting electron-hole excitations.  

Our results further show that
the magnetic excitations in the intermediate coupling regime are qualitatively
similar to those found at strong coupling, for all dopings investigated. The same
observation does not hold for the charge excitations. Indeed, when the Hubbard
repulsion is of the order of the bonding/anti-bonding bandwidth, smaller
hole-doping concentrations are sufficient to transfer much of the charge
spectral weight to low energy intraband excitations. We can naively explain
these observations by noting that Hubbard $U$ interactions \emph{directly}
affect the charge degrees of freedom while it only indirectly affects the spin
degrees of freedom of the system via the antiferromagnetic exchange
interaction. Moreover, RPA+FLEX approximation is
expected to fail at large $U$.

Our results are also of direct relevance to inelastic neutron
scattering (INS) and resonant inelastic x-ray scattering (RIXS) experiments on
two-leg ladder
cuprates~\cite{Ishii2007,Wray2007,notbohm2007,higashiya2008significance,re:schlappa2009,roth2010plasmons}.
The $S(\textbf{q},\omega)$ spectra in the undoped case at strong coupling are in good
qualitative agreement with available experimental INS data, showing one-triplon
and two-triplon excitations~\cite{notbohm2007,re:schlappa2009}.  We believe
that the dispersive incommensurate features found in our magnetic excitation
spectra at finite hole doping may be detectable by INS in two-leg ladders
telephone number compounds (La,Sr,Ca)$_{14}$Cu$_{24}$O$_{41}$. 

Concerning the dynamical charge structure factors, our DMRG results show good
qualitative agreement with a recent RIXS experiment on the hole-doped two-leg
ladder cuprate compounds (La,Sr,Ca)$_{14}$Cu$_{24}$O$_{41}$~\cite{Ishii2007}.
In this experimental work, two kinds of excitations appear in the RIXS spectra.
One is attributed to an interband excitation across the Mott gap, observed at
2--4~eV with a dispersion relation that is independent of the hole-doping
concentration of the ladder.  The second excitation appears as a continuum
below the Mott gap energy 2~eV when holes are doped, and its intensity is found
to be proportional to the hole-doping concentration.  We observe this same
qualitative behavior in our $N(\textbf{q},\omega)$ spectra in the strong couplig regime
in both the $q_\text{rung}=0$ and $\pi$ components for small hole-doping up to
$10\%$ [see Fig~\ref{fig6}(g)-(h)]. Moreover, the spectral weight of $N(\textbf{q},\omega)$ is redistributed to a low energy intraband excitations in the
overdoped regime [see spectra for $U/t=6$ and $n=0.666$, corresponding to
$33\%$ hole-doping in Fig.~\ref{fig6}(i)]. In our $N(\textbf{q},\omega)$ spectra, we
found that most of the charge spectral weight appears in a low energy band,
that is quite dispersive across the Brillouin zone in contrast to the results
shown in Ref.~\cite{Ishii2007}.
  
\begin{acknowledgments}
A. N., T. M. and E. D. were supported by the US Department
of Energy (DOE), Office of Basic Energy Sciences
(BES), Materials Sciences and Engineering Division. 
G. A., T. M. and S. J. were supported by the Scientific Discovery through Advanced Computing (SciDAC) program funded by U.S. Department of 
Energy, Office of Science, Advanced Scientific Computing Research and Basic Energy Sciences, Division of Materials Sciences and Engineering. N. P. was supported by the National
Science Foundation Grant No. DMR-1404375. 
Numerical simulations were performed at the Center for Nanophase Materials Sciences, 
which is a DOE Office of Science User Facility.
\end{acknowledgments}

\bibliography{thesis}

\end{document}